\documentclass{emulateapj}	
\usepackage{times}
\usepackage{graphicx}

\slugcomment{To appear in ApJ}

\def\lsim{\mathrel{\rlap{\lower4pt\hbox{\hskip1pt$\sim$}}
    \raise1pt\hbox{$<$}}}                
\def\gsim{\mathrel{\rlap{\lower4pt\hbox{\hskip1pt$\sim$}}
    \raise1pt\hbox{$>$}}}                

\shorttitle{PSR J1311$-$3430 Spectroscopy}
\shortauthors{Romani et al.}

\begin{document}

\title{A Spectroscopic Study of the Extreme Black Widow PSR J1311$-$3430}

\author{Roger W. Romani\altaffilmark{1}, Alexei V. Filippenko\altaffilmark{2}, and
S. Bradley Cenko\altaffilmark{3,4}
}

\altaffiltext{1}{Department of Physics, Stanford University, Stanford, CA 94305-4060,
 USA; rwr@astro.stanford.edu}
\altaffiltext{2}{Department of Astronomy, University of California, Berkeley, CA
 94720-3411, USA}
\altaffiltext{3}{Astrophysics Science Division, NASA/Goddard Space Flight Center, 
MC 661, Greenbelt, MD 20771, USA}
\altaffiltext{4}{Joint Space-Science Institute, University of Maryland, College 
Park, MD 20742, USA}

\begin{abstract}

	We report on a series of spectroscopic observations of PSR J1311$-$3430, an extreme 
black-widow gamma-ray pulsar with a helium-star companion. In a previous study we estimated the
neutron star mass as $M_{\rm NS}= 2.68 \pm 0.14\,{\rm M}_\odot$ (statistical error), based on limited 
spectroscopy and a basic (direct heating) light curve model; however, much larger model-dependent
systematics dominate the mass uncertainty. Our new spectroscopy reveals a range of complex source behavior.
The variable He~I companion wind emission lines 
can dominate broad-band photometry, especially in red filters or near minimum brightness, and the wind flux
should complete companion evaporation in a spin-down time. The heated companion face 
also undergoes dramatic flares, reaching $\sim$\,40,000\,K over $\sim 20$\% of the star; this is likely powered
by a magnetic field generated in the companion. The companion center-of-light radial velocity is now well measured
with $K_{\rm CoL} = 615.4 \pm 5.1$\,km\,s$^{-1}$. We detect non-sinusoidal velocity components due to the heated face
flux distribution. Using our spectra to excise flares and wind lines, we generate substantially improved
light curves for companion continuum fitting. We show that the inferred inclination and neutron star mass, however,
remain sensitive to the poorly constrained heating pattern. The neutron star's mass, $M_{\rm NS}$, is likely 
less than the direct heating value and could range as low as 1.8\,M$_\odot$ for extreme equatorial 
heating concentration. While
we cannot yet pin down $M_{\rm NS}$, our data imply that an intrabinary shock reprocesses the 
pulsar emission and heats the companion. Improved spectra and, especially, models that include such shock 
heating are needed for precise parameter measurement.
\end{abstract}


\section{Introduction}

PSR J1311$-$3430 (hereafter J1311) is a ``black widow" millisecond pulsar \citep[MSP;][]{pet12} 
with spin-down luminosity ${\dot E} = 5 \times 10^{34}\,I_{45}$\,erg\,s$^{-1}$ (for a neutron star
moment of inertia $10^{45}\, I_{45}$\,g\,cm${^2}$) in a $P_B= 93.8$\,min
orbit \citep{r12}, the shortest of any confirmed spin-powered pulsar.\footnote{J1653$-$0158 \citep{ret14} may be a similar system with even smaller $P_b=74.8$\,min, but it has not yet had a pulse detection.} J1311 is
particularly interesting, as photometric and spectral studies \citep[][hereafter R12]{ret12}
show that it has a strong, violently variable evaporative wind driven from a $M_2 = 0.01\,{\rm M}_\odot$
He companion and that the neutron star is particularly massive. In R12 we found that
with standard assumptions for heating of the companion by the pulsar, the light-curve 
modeling gave $M_1 = 2.68 \pm 0.14\,{\rm M}_\odot$ (statistical errors),
although with different assumptions for asymmetric heating one could obtain
masses $\sim 2.1$--2.9\,M$_\odot$ (systematics range). This system
promises to reveal much about close binary evolution \citep{BDH13,vHet12},
the efficacy of the companion evaporation channel for producing isolated MSPs, and
(possibly) fundamental constraints on the equation of state of matter at supernuclear densities.

	Measurement of the orbital motion, heating, and wind stripping of the companion is the 
key to such studies.  To this end, we have made optical spectroscopic observations
at several epochs.  Here we summarize the measurements, describe the new phenomena discovered,
introduce new models of the system, fit for the component masses, and comment on the results.

\begin{figure}[b!!]
\vskip 5.4truecm
\includegraphics{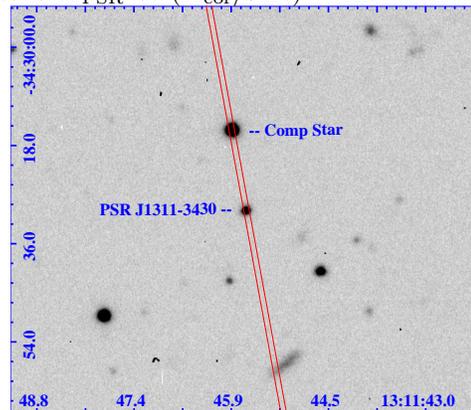}
\begin{center}
\caption{\label{finder} 
120\,s $g^\prime$ SOAR/SOI image of PSR J1311$-$3430 at maximum 
$\phi_B \approx 0.75$, showing the pulsar, the comparison star, and the
1$^{\prime\prime}$-wide slit position.
}
\end{center}
\vskip -0.7truecm
\end{figure}

\section{Spectroscopic Campaigns}

	In R12, we discussed photometric evidence for substantial optical flaring activity of J1311.
These flares appear to have a red spectral energy distribution (SED), 
but further study of this behavior requires
spectroscopic characterization of the variable component. Also, multi-orbit photometry shows
variable asymmetry about the optical maximum. This limits the reliability of model fitting (with
simple symmetric pulsar heating) in determining the system inclination $i$ and the 
correction factor $K_{\rm cor} = K_{\rm CoM}/K_{\rm CoL}$, where $K_{\rm CoL}$ is
the observed radial-velocity amplitude from spectral features on the heated face of the companion
(the center of light; CoL) and $ K_{\rm CoM}$ is the true radial-velocity amplitude of the companion center of mass (CoM).
These factors are critical, since for a given observed radial-velocity amplitude the pulsar 
mass is $M_{\rm PSR} \propto (K_{\rm cor}/{\rm sin}\, i)^3$.
The detailed behavior of the photospheric absorption lines can probe the
surface-temperature distribution, and emission-line components can probe the evaporating wind.

	Thus, intensive spectroscopic campaigns offer the best hope for further understanding
this important system. We report here on our efforts to date with the Keck-I/Low Resolution
Imaging Spectrometer (LRIS; Oke et al. 1995) and
the Gemini/GMOS-S \citep{het05}  systems. For both, we sought long-slit multi-orbit coverage with high resolution
and sensitivity. By placing the slit at a position angle (PA) of $10^\circ$, we were able to simultaneously
observe the pulsar and a  $r^\prime = 18.05$\,mag G-type comparison star 15$^{\prime\prime}$ from
J1311 (Figure \ref{finder}). The latter allowed us to stably cross-calibrate the flux and wavelength, 
permitting coherent analysis of both datasets.  With Keck-I we suffered from poor weather, 
with only half a night of the two scheduled nights delivering J1311 spectra, under
marginal conditions. Nevertheless, we obtained three orbits of continuous coverage.
The GMOS-S observations were queue scheduled and so all data were usable, with three visits
covering slightly more than one orbit each. We describe these observations in detail,
along with several short supplementary spectral sequences which give context to the variable 
behavior, especially near binary optical maximum.

\subsection{March 2013 LRIS Spectroscopy}

	We observed J1311 with Keck-I/LRIS on 2013 March 11 (MJD 56362.429--56362.626),
using the 1$^{\prime\prime}$-wide slit,
a dichroic splitter at 5600\,\AA, the 600/4000 grism in the blue camera, and the 400/8500 grating
in the red camera. Table 1 lists the wavelength coverage and resolution for all observations.
Our goal was to obtain high signal-to-noise ratio (S/N) spectra at all orbital phases.
With a high southern declination,
observations from Mauna Kea were perforce at moderate airmass (${\rm sec} z$ = 1.7--2.5). The
seeing was poor (mostly $>1.3^{\prime\prime}$) and the transparency variable. Each exposure was
300\,s in duration; this was a compromise between improving the S/N but not having too much
orbital smearing.

\begin{figure*}[t!!]
\vskip 12.2truecm
\includegraphics{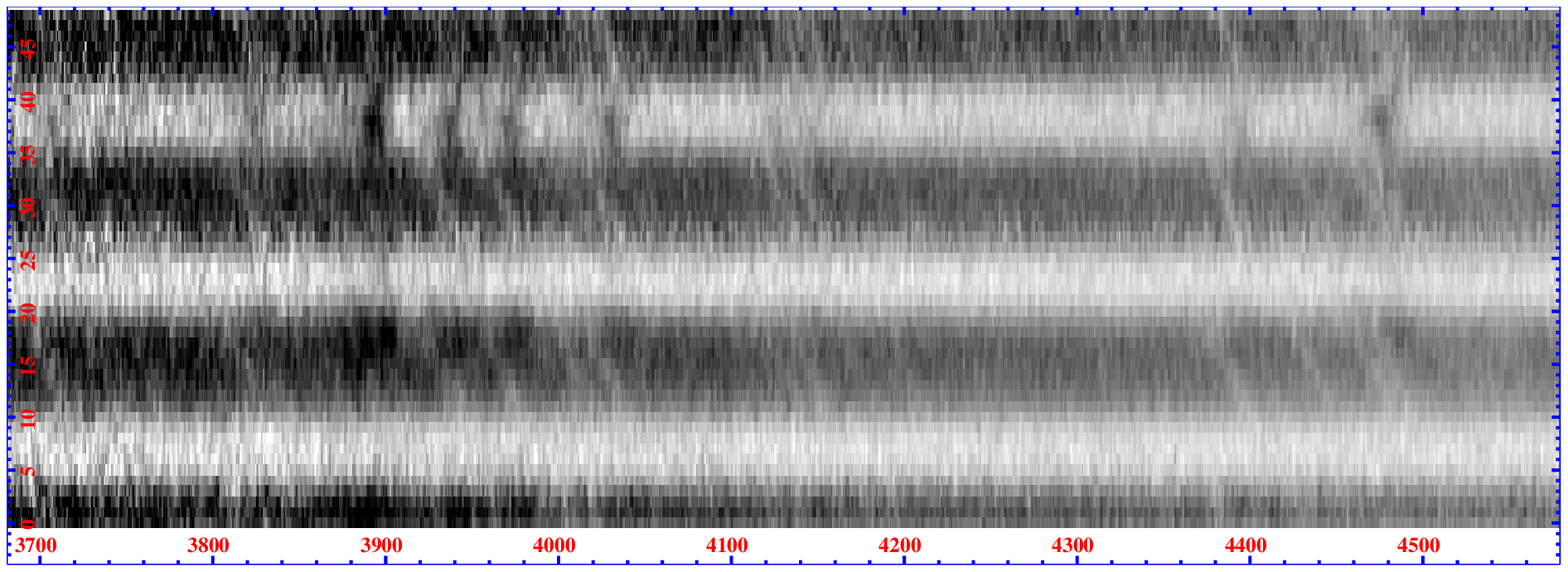}
\includegraphics{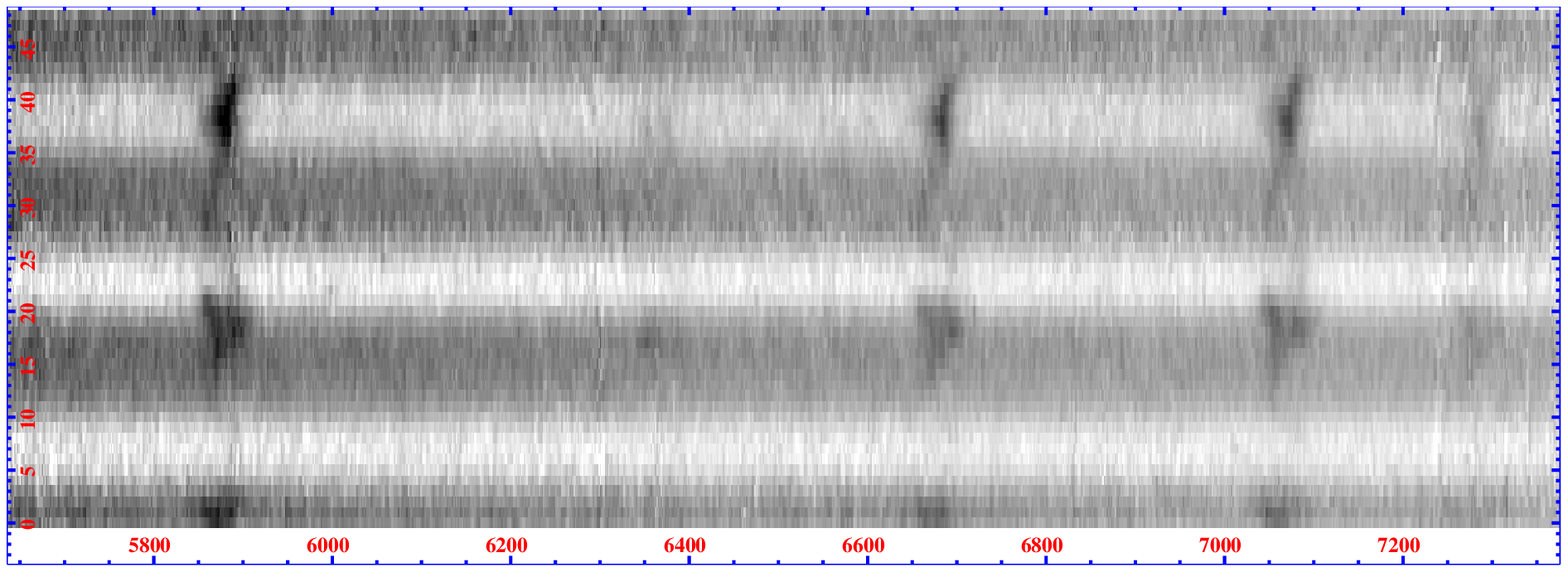}
\begin{center}
\caption{\label{trailedspec} 
Trailed spectra from the 2013 March Keck/LRIS campaign. Fluxes have been normalized to
remove variations in the simultaneously observed comparison star. 
Upper panel: Blue side, showing Doppler-shifting absorption features from
the companion photosphere. Lower panel: Red side, showing variable He~I emission features from
the companion wind. The scale is inverted: darker indicates more flux.
}
\end{center}
\vskip -0.7truecm
\end{figure*}

   Standard processing 
and optimal extraction were applied; fluxes were calibrated against the red-side and blue-side 
standard stars BD+26$^\circ$2606 and BD+28$^\circ$4211, respectively.  The comparison-star flux 
varied by a factor of 1.7 during this observation sequence, largely owing to variable 
seeing-induced slit losses.  After normalizing the flux, as monitored by the comparison star,
we show (Figure \ref{trailedspec}) ``trailed spectra" for the central portions of the blue and 
red sides. 

	The 49 spectra cover just over three orbits ($\phi_B = -0.18$ to 2.85),
with the midpoint of the barycentric arrival time of each exposure
referenced to the ephemeris of \citet{pet12}. Here we set $\phi_B = 0$ at the first pulsar ascending
node after the start of observations. For convenience, we will follow \citet{bet13} in referring
to phase near optical maximum (pulsar superior conjunction, $\phi_B = 0.75$) as ``Day'' and
phases viewing the companion's unheated face ($\phi_B = 0.25$) as ``Night.'' ``Dawn'' and ``Dusk''
are at phases $\phi_B=0.5$ and  $\phi_B=0.0$, respectively.

	Several features are evident in these trailed negative (darker means more flux) spectra.
First, we see the continuum fading and brightening from spectrum 1 (bottom) to 49 (top).
The continuum modulation is stronger in the blue than in the red, as expected
from the high temperature of the companion's heated Day face. Superposed on this 
continuum are Doppler-shifted absorption features (R12). We also see 
broad and variable emission lines, strongest in the Dusk and Night phases. 
While little emission is visible during the first orbital minimum (spectra 5--10), the emission
lines are very strong in the third Night (spectra 35--40), and indeed have such
large equivalent width that they dominate in the red. Since the companion is very H-poor
(R12), both the photosphere and the wind trail driven from it are dominated by He~I.

\subsection{2013 Gemini Spectroscopy}

	The Keck observations have good sensitivity and broad wavelength coverage, but
lack the resolution to detect the line-shape distortions
imposed by nonuniform companion heating. To search for such effects, we
observed J1311 with Gemini, using the GMOS-S spectrograph in a novel mode to maximize
resolution in the blue part of the spectrum which displays many photospheric He~I and 
metal lines. These observations used the R831 grating in second order through a
0.75$^{\prime\prime}$ slit, which covered 3950--5100\,\AA\ at
$R \equiv \lambda/\delta \lambda \approx 5000$, $\sim 60\,{\rm km\,s^{-1}}$ resolution. This configuration required
an order-blocking filter. The best available filter, $g^\prime$, unfortunately has a
4000\,\AA\ cutoff which decreases sensitivity to Ca~H; moreover, the protected silver coating
on the primary mirror greatly limits its effective area below 4000\,\AA.
To minimize the orbital velocity smearing, we restricted exposures to 175\,s (i.e., 
maximum smearing of $60\,{\rm km\,s^{-1}}$ at quadrature), obtaining 38 consecutive 
exposures during each of three visits to the source.  
Each visit started near an optical maximum and covered $\Delta \phi_B = 1.27$
in phase. The spectrum orbital phase is set by the exposure midpoint 
at the solar-system barycenter. 

	The data were processed using the Gemini IRAF package. Fluxes were established using
observations of the A0 standard CD~32$-$9927. Comparison-lamp exposures were only taken once per visit,
but, as for the LRIS spectroscopy, we monitored velocity stability and slit losses using the
nearby comparison star. This also facilitated comparison between observing runs.
The first visit, on 2013 May 6 (MJD 56418.15328--56418.23601), featured excellent 
$\sim 0.5^{\prime\prime}$ seeing and high throughput, while the next
two visits, 2013 June 3 and 5 (MJD 56446.04081--56446.12281 and 56448.05900--56448.14101),
had more typical $\sim 0.8^{\prime\prime}$ image quality and lower S/N. 

\begin{figure}[t!!]
\vskip 9.3truecm
\includegraphics{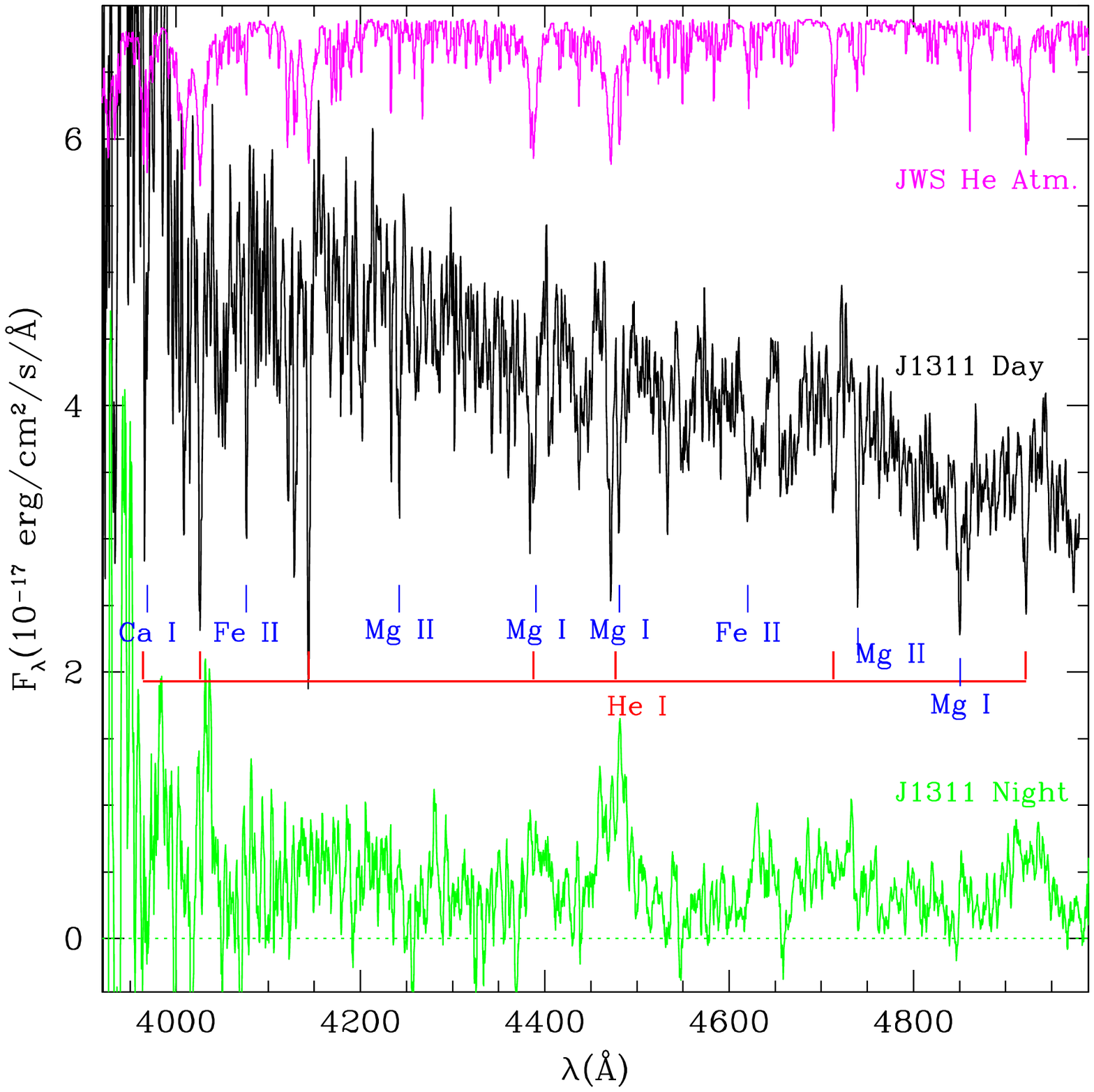}
\begin{center}
\caption{\label{GemSpec} 
Average spectra near maximum light ($\phi_B = 0.65$--0.85, black curve) and 
minimum light ($\phi_B = 0.15$--0.35, green curve)
from the GMOS-S data. A comparison He-dominated model atmosphere (magenta curve) shows that many
strong, narrow absorption features are present in the maximum-light ``Day" spectrum; some
of the strongest lines are marked. Broad emission is apparent at minimum light. 
}
\end{center}
\vskip -0.7truecm
\end{figure}

In Figure \ref{GemSpec} we show the average spectrum from maximum light ($\phi_B = 0.65$--0.85,
Day, all GMOS-S orbits) and near minimum light ($\phi_B = 0.15$--0.35, Night, all GMOS-S orbits), Doppler corrected
to rest using the radial-velocity solution (see below). For comparison, the upper trace shows
a He-dominated model atmosphere with $T_{\rm eff} =$ 12,000\,K and ${\rm log}\,g = 4.5$ \citep{jwp01}.
In the spectrum from J1311 maximum, many narrow absorption features of He~I and low-excitation
metal lines are seen. In contrast, the average spectrum at minimum shows broad emission
features, with He~I and Mg~I most prominent. Unlike the red LRIS spectra, these do not
strongly dominate the 4000--5000\,\AA\, continuum, so we cannot independently measure the wind
emission-line profile and variation. In the trailed spectra this line emission is most obvious
around $\phi_B \approx 0$ (centered on ``Dusk'' phases) during the first visit. It is 
weaker but present in the rest of the GMOS-S data.

\subsection{Other Spectroscopy}

	Since the wind features sometimes show dramatic changes from one orbit to the next, 
we also describe several other observations of J1311, designed to check the system's status. 
In R12, we discussed our initial six consecutive 300\,s LRIS exposures 
covering $\phi_B = 0.56$--0.93 on 2012 May 17 (MJD 56064.306--56064.331). These data, obtained 
with the same configuration as on 2013 March 11,
established the companion He dominance and provided 
an estimate $K_{\rm obs}=609.5 \pm 7.5$\,km\,s$^{-1}$ for a sinusoid fitting the CoL radial velocity.
At this epoch J1311 appeared to be in quiescence, with no emission lines.

	We also observed J1311 on 2013 April 13 (MJD 56390.424--56390.433)
for $3 \times 300$\,s near optical maximum ($\phi_B  = 0.75$) using the 
DEep Imaging Multi-Object Spectrograph (DEIMOS; Faber et al. 2003) on Keck-II.
DEIMOS provided coverage of 4450--9630\,\AA, with the dichroic set at 6990\,\AA.
Standard extraction, flux calibration, and cleaning were applied. Observations were made at the
parallactic angle \citep{fil82} and did not include the comparison star.

\begin{figure}[b!!]
\vskip 9.1truecm
\includegraphics{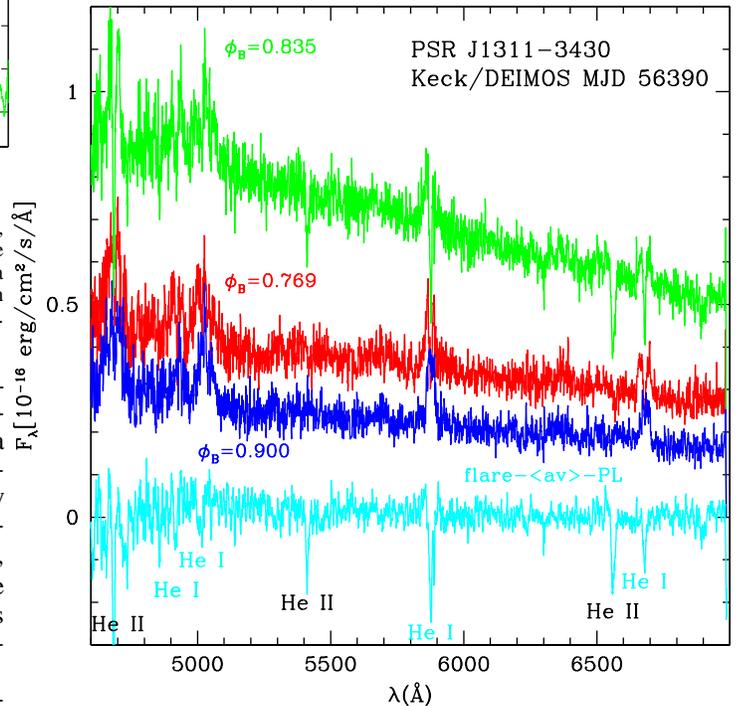}
\begin{center}
\caption{\label{Deimos} 
Sequential spectra from the blue side of the Keck-II/DEIMOS observations. 
He~I emission from the wind is present in all spectra. The flux doubles
during the second exposure (green curve), and the residual spectrum (cyan, at bottom) shows the
absorption features after subtracting the continuum 
power law: He~II absorption is strong. This flare component is only weakly
present in the first spectrum (red curve), and no He~II absorption is apparent during 
the third spectrum (blue curve) 600\,s later.
}
\end{center}
\vskip -0.7truecm
\end{figure}

	The companion was in a particularly bright state with a dramatic flare during 
the second spectrum (centered at $\phi_B=0.835$). During this flare the continuum
flux increased by a factor of $\sim 2.5$.  He~II absorption
appears quite strong, particularly in the second spectrum.  Figure  \ref{Deimos} shows
the dramatic continuum variability over $\sim 300$\,s at this epoch. Note that
the difference spectrum at bottom shows that the emission-line variations, if any,
are quite small. We thus infer that this heating is applied below the photosphere, allowing 
relatively narrow absorption features to appear. The broader superimposed emission
lines arise from a different location.

Finally, we returned to J1311 with Keck-I/LRIS on 2013 May 10 (MJD 56422.347--56422.355) for
three additional 300\,s exposures, this time covering phases
$\phi_B=0.012$--0.139. The source appears to have returned to quiescence with
faint line emission.

	In Table 1 we summarize these various spectroscopic campaigns. To give
some idea of the state of the wind line emission, we list the flux of the
He~I $\lambda$6678 line (we measure the He~I $\lambda$4471 line for GMOS-S spectra),
after averaging over the full visit. Of course, as apparent from Figure 
\ref{trailedspec}, the line strength is highly variable over even a single orbit,
so these fluxes should be considered only a rough characterization of the
wind activity. For example, although the line emission was not detected during
the average of the second and third GMOS-S visits, we can detect He~I emission
during the Night $(\phi_B \approx 0.25$) portion of the orbit (Figure \ref{GemSpec}).

\begin{figure}[b!!]
\vskip 9.2truecm
\includegraphics{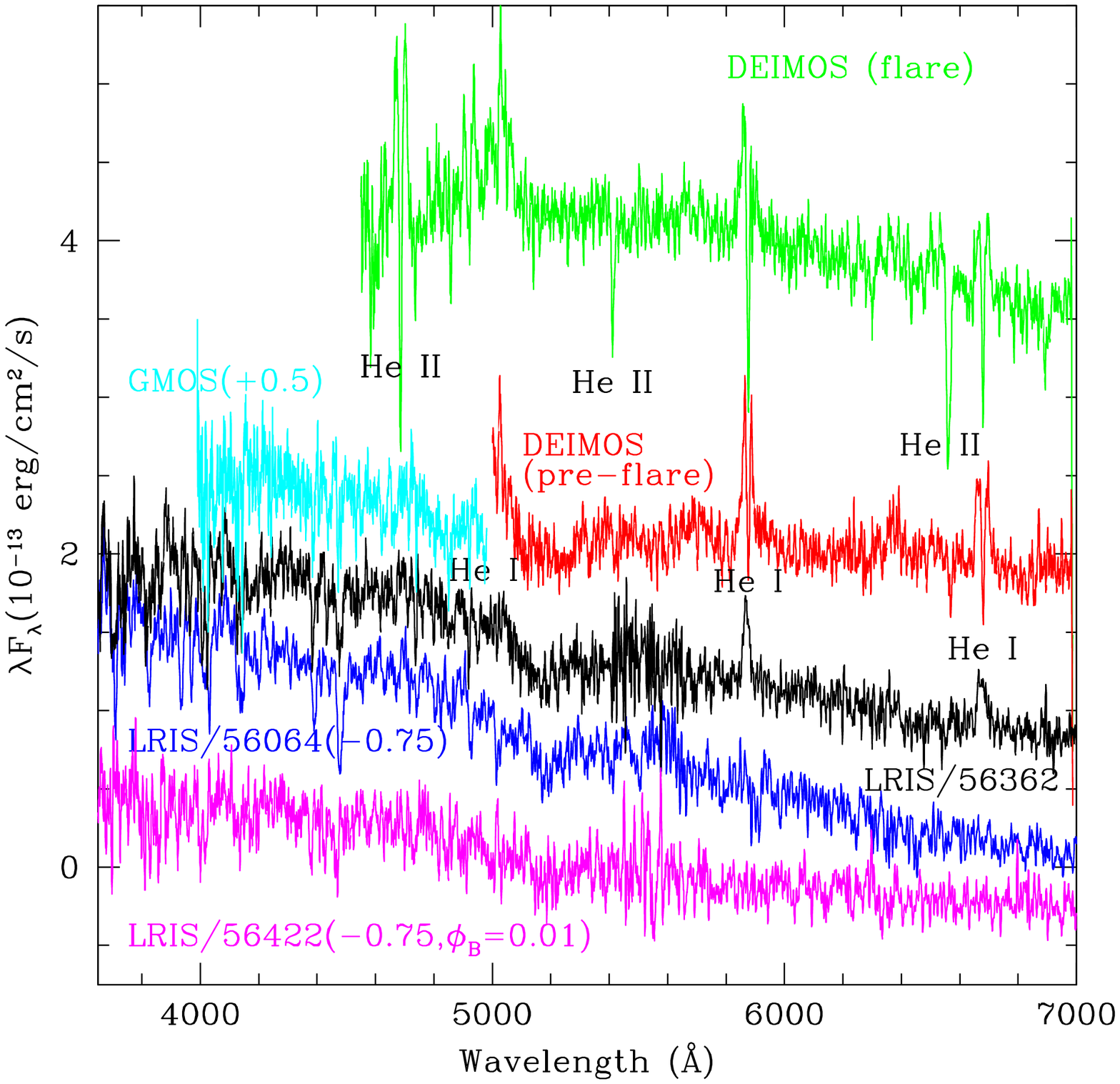}
\begin{center}
\caption{\label{SpecComp}
Spectra near optical maximum ($\phi_B = 0.65$--0.85) from our five observing 
campaigns (grouping the three GMOS-S visits). The GMOS-S (cyan) and LRIS/56064 (blue) 
spectra have very similar flux to that of our long LRIS/56362 campaign (black) 
and are offset for readability. The LRIS/56422  (magenta) is also
offset by $-0.75$, but is at ``Dusk" phase and so is intrinsically fainter.
The other spectra have flux
as plotted. Note the dramatic continuum increase during the DEIMOS/56390 flare (green),
during which He~II absorption is prominent. Also note the strong and variable He~I 
emission in the red. 
}
\end{center}
\vskip -0.7truecm
\end{figure}

Figure \ref{SpecComp} compares the $\lambda F_\lambda$ spectra near 
$\phi_B=0.75$ from these five datasets. The GMOS observation and the 
LRIS/56064 spectrum (here and elsewhere, the label refers to MJD) have fluxes 
similar to that of LRIS/56362, but have been 
offset for readability. We also offset the LRIS/56422 spectrum by $-0.75$;
note that, unlike the other spectra in this figure, this is at quadrature
($\phi_B=0.01$) and thus is intrinsically fainter.
Two DEIMOS/56390 spectra near maximum light are also shown: phase
$\phi_B=0.77$ before the main flare (only $\lambda > 5000$\,\AA\ plotted to avoid
overlap) and the flare peak at $\phi_B=0.84$. By $\phi_B=0.90$, J1311 returns to
the quiescence level. All quiescent spectra have similar continuum fluxes, but
the He~I lines vary greatly: absent on 56064, strong on 56362, and a factor of 
$\sim 2$ stronger on 56390.

\begin{deluxetable*}{llllrrl}[t!!!!!]
\tablecaption{\label{Journal} PSR J1311$-$3430 Spectroscopic Campaigns}
\tablehead{
\colhead{Date} & \colhead{Tel./Instr.}& \colhead{Range} & \colhead{Res} &\colhead{Exp.} & \colhead{Orb. Phase} 
&\colhead{$\langle$He~I 6678$\rangle$}\cr
\colhead{(MJD)} & \colhead{}& \colhead{(\AA)} & \colhead{(\AA)} &\colhead{($N \times s$)} & \colhead{($\phi_B$)} & \colhead{
($10^{-18}\,{\rm erg\,cm^{-2}\,s^{-1}}$)}
}
\startdata
56064\tablenotemark{a} &Keck-I/LRIS    & 3100--10,500 & 4,7\tablenotemark{b} & $6\times 300$  & 0.56--0.93 &$< 3$\cr
56362 &Keck-I/LRIS    & 3100--10,500 & 4,7\tablenotemark{b} & $49\times 300$ & $-0.18$--2.85 & 200. \cr
56390 &Keck-II/DEIMOS & 4450--9630 & 4.6 & $3\times 300$   & 0.77--0.90 & 390.\cr
56418 &Gemini/GMOS-S  & 3950--5100 & 0.9 & $38\times 175$  & 0.58--1.85 & 28.\tablenotemark{c} \cr
56446 &Gemini/GMOS-S  & 3950--5100 & 0.9 & $38\times 175$  & $-0.16$--1.10 & $<10$\tablenotemark{c}\cr
56448 &Gemini/GMOS-S  & 3950--5100 & 0.9 & $38\times 175$  & $-0.17$--1.09 & $<8$\tablenotemark{c} \cr
56422 &Keck-I/LRIS    & 3100--10,500 & 4,7\tablenotemark{b} & $3\times 300$  & 0.01--0.14 & 9.0\cr
\enddata
\tablenotetext{a}{Previously described by \citet{ret12}; all other observations presented here for the first time.}
\tablenotetext{b}{Blue-side resolution, Red-side resolution.}
\tablenotetext{c}{No red-side coverage; estimate from He~I $\lambda$4471.}
\end{deluxetable*}

\section{The Wind Kinematics}

\begin{figure}[t!!]
\vskip 4.7truecm
\includegraphics{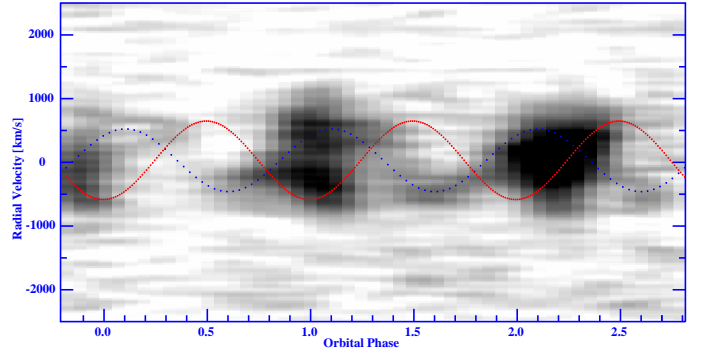}
\begin{center}
\caption{\label{EmLines} 
Velocity profile of the He~I emission from the companion wind on MJD 56362.
For reference, the red curve shows the radial velocity of the
simple sinusoidal fit to the absorption features, marking the CoL
velocity of the heated surface of the companion star. The emission lines
do not track this motion, but a likely sinusoidal component is shown by the dotted
line. Over three periods are shown, with dark colors marking bright line emission. 
}
\end{center}
\vskip -0.7truecm
\end{figure}

	The broad He~I emission, which must arise in the companion wind, is clearly
episodic and variable. Our best measurement of this emission is in the March 2013 Keck/LRIS data
(MJD 56362).  Since the lines appear optically thin with relatively 
constant line ratios, we have formed a wind-velocity profile by taking the weighted, continuum-subtracted
average of the three strongest He~I lines in the red: 5875.61\,\AA, 6678.15\,\AA,
and 7065.18\,\AA. The result (Figure \ref{EmLines}) shows the flaring, varying emission-line 
profile. Scaled versions of this line profile provide good matches to the emission-line 
structure in the blue, as well.

In Figure \ref{EmLines},
although the emission is patchy, there is approximate orbital periodicity, with one component
following the dotted sinusoid. Note that the emission other than this sinusoidal component seems to generally
lie redward of the companion CoM velocity.  We conclude that some preferred vectors of emission 
in the rotating system exist, but that the emission is intermittent along these vectors.

\begin{figure}[t!!]
\vskip 9.5truecm
\includegraphics{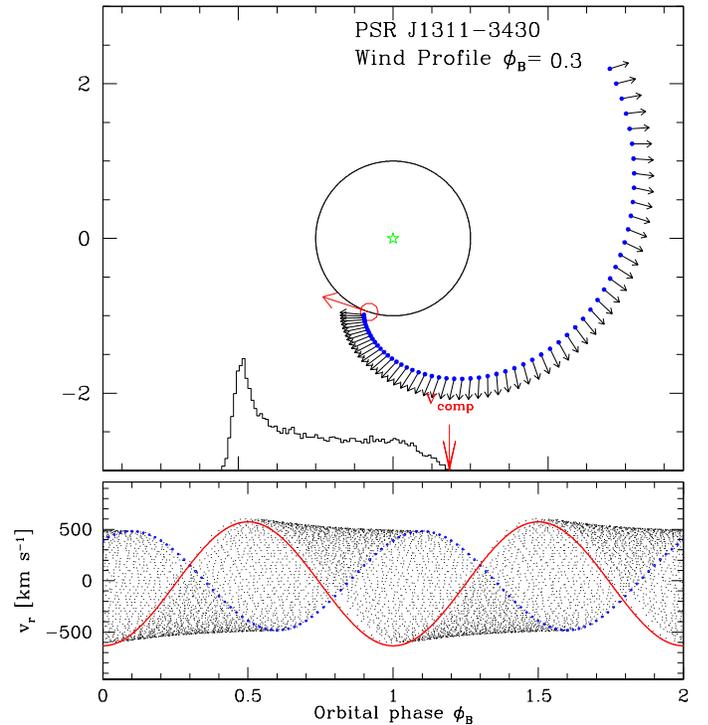}
\begin{center}
\caption{\label{WindMod} 
A simple ballistic model for the cometary trail emission. Mass is lost from the companion
and accelerated radially at $3/4\, g_{\rm PSR}$, via the pulsar spin-down flux. One phase, with
the trail (out to a lag $\Delta \phi_B=0.7$), along with the line profile (histogram) for uniform illumination,
is shown at top. The red arrows show the companion velocity in space and at the red edge of
the line profile, for $\phi_B=0.3$.
Two periods of the trail velocity range along with the companion velocity and
the sinusoidal component curves (curves exactly as in Figure \ref{EmLines}) are shown in the lower panel.
}
\end{center}
\vskip -0.7truecm
\end{figure}

	Figure \ref{WindMod} show a schematic model illustrating a possible
kinematic origin for the wind line structure. Here we compute the cometary wind trail for
matter stripped at low velocity (in the corotating frame) from the companion and then accelerated
by pulsar radiation pressure; the only free parameter in this model is $a_{\rm PSR}/g_{\rm PSR}$,
the ratio of wind acceleration to gravity. A similar model with the addition of a finite launch speed 
of 2/3\,$v_{\rm orb}$ does a good job of defining the radio eclipse envelope for the original black-widow
PSR B1957+20 \citep{rest89}.  Phase $\phi_B=0.3$ is shown in the upper panel, with the arrows and
histogram indicating the possible radial velocities along this wind trail; a patch of He-containing
gas will provide an emission line within this envelope. The lower panel illustrates the range of velocities
as a function of phase. This figure can be compared with Figure \ref{EmLines}; we reproduce the two curves
from that figure to guide the comparison. The wind trail plotted fades after $\Delta \phi_B=0.7$; this
boundary matches well in both amplitude and phase the quasi-sinusoid traced in Figure \ref{EmLines} 
by the dotted curve. If the wind emission is relatively strong at lag $\Delta \phi_B=0.7$,
patchy at smaller $\Delta \phi_B$, but stronger at $\phi_B \approx 0.75$--0.25, this pattern
provides a reasonable match to the data. Many orbits would need to be monitored to
determine if this represents a dominant persistent emission pattern.

	This kinematic model is attractive, but the present data do not allow us to determine the 
physical origin of the emission-line velocities; of course, any simple sinusoid can be reproduced 
by a rotating vector fixed in the binary frame, for example at the companion star. The dotted 
curve shown can be produced by a vector (i.e., a jet) from the companion with $v_j\approx 1.3 v_{\rm orb)}$ in the 
star's rest frame, directed in the orbital plane $\sim 60^\circ$ ahead of the line
of centers. The physical nature of the pulsar flux driving the tail is also unclear. With ${\dot E}=
5 \times 10^{34}\,I_{45}\,{\rm erg\,s^{-1}}$, $a_{\rm PSR}/g_{\rm PSR}=3/4$ requires a mean wind cross section of 
$\langle \sigma/m \rangle \approx 1.5 \times 10^3\, {\rm cm^2\,g^{-1}}$. This is about right for 
a ionized pair plasma, but high for a baryonic wind. It is also interesting to consider what
this wind tells us about pulsar evaporation. The momentum density of the pulsar wind 
is $40\, I_{45}\, {\rm g\,cm^{-1}\,s^{-2}}$ at the orbital radius. To ram-pressure balance this wind along
the line of centers requires a companion wind density 
$\rho_{\rm W} = 1.0 \times 10^{-14}\, I_{45} v_{600}^{-2}\,{\rm g\,cm^{-3}}$
and a total mass flux ${\dot m}_{\rm W} = \pi R_\ast^2 \rho_{\rm W} v_{\rm W} \approx
6 \times 10^{13}\, I_{45} v_{600}^{-1}\,{\rm g\,s^{-1}}$, for a wind speed $600\, v_{600}{\rm km\,s^{-1}}$ 
comparable to the orbital speed
and wind only from the heated face of the companion. Note that this gives an evaporation timescale
$\tau \approx m_2 / {\dot m}_{\rm W} \approx 10\, m_{-2} v_{600}/I_{45}$\ Gyr,
for $m_2=10^{-2}\,m_{-2}\,{\rm M}_\odot$. Although $m_2$ 
was larger when accretion stopped and evaporation began, ${\dot E}$ was higher in the past
so the companion may be completely ablated in less than a spin-down time,
especially for a slow wind or $I_{45} >1$.  

\section{The Photosphere Flare Variability}

	Figures \ref{Deimos} and \ref{SpecComp} show that in addition to the emission-line variability,
there can be epochs of true continuum increase. The dramatic $i^\prime$ flare shown in Figure 1 of \citet{r12},
which erupts to 6 times the normal flux at $\phi_B\approx 0.8$, is likely a similar event.  Figure
\ref{Deimos} gives an important clue to the nature of these variations. The difference spectrum
shows that the flare does not affect the emission-line fluxes, but produces 
strong absorption lines. The He~II absorption dramatically increases (including a very 
strong, unusual detection of He~II $\lambda$6560, generally confused with the absent H$\alpha$ line).
We infer that these flares must be caused by impulsive deep heating well below the photosphere.
However, since the photometric flares and spectral flares have decay times of $\sim 300$\,s,
if we assume a sound speed $\sim 8\, T_{4.5}^{1/2}{\rm km\,s^{-1}}$ in the $\sim$\,30,000\,K
He gas seen during these bursts, we infer a characteristic depth of $\sim 2\times 10^8$\,cm
($\sim 0.03\, R_\ast$). 

By measuring the He~II/He~I line-intensity ratios in the flare spectrum,
compared with the ratios in model spectra of \citet{jwp01},
we find an effective flare temperature of $T_{\rm Fl} =$ 39,000 $\pm$ 1000\,K.
The continuum flux increase from quiescence is a factor of 2.2 at 5500\,\AA\ 
assuming Planckian continua in this range. This implies a flare-emitting
area 0.27 times that of the quiescent star. Thus, we infer that a large fraction
of the star brightened by a factor of 8.3 during the eruption. At the estimated
(from the dispersion measure)
distance of 1.4\,kpc, the equivalent radius of the flare area is 0.05\,R$_\odot$,
which is $\sim 0.5\,R_\ast$, in good agreement with the spectral estimate. 

The average thermal luminosity of this burst is 
$\sim 5 \times 10^{33}\,{\rm erg\,s^{-1}}$. This is $\sim 80$ times the pulsar
flux hitting the companion, so we infer that these outbursts represent stored
energy released deep below the photosphere. The natural candidate energy source
is a magnetic field, with characteristic magnitude $\langle B \rangle \approx 10$\,kG
required to power the flare.
The origin of such a field is unclear. Typically stars with $T_{\rm eff} =$ 12,000\,K
are radiative throughout. However, here we expect that high-energy particles from the
pulsar are deeply heating the photosphere, with (for example) TeV $e^\pm$ depositing their energy
at $\Sigma \approx 300\,{\rm g\,cm^{-2}}$.  While this does not in itself induce convection,
we expect latitudinal flow because of the large temperature gradients from the differential heating.
If convective, given the short effective spin period for
the tidally locked companion, we have the ingredients needed for a robust
dynamo, large magnetic fields, and energetic flare events.

\section{Improved Quiescence Light Curves from Spectra}

	The large equivalent width of the variable He~I lines can
certainly affect the broad-band fluxes, especially at binary minimum and for red colors.
This is likely the origin of the red ``flaring events" noted by R12 in simultaneous 
multi-band GROND light curves. We can use our spectra to partly correct for this effect,
synthesizing light curves from the continuum. To do this, we integrate our normalized 
spectra over the SDSS $g^\prime r^\prime i^\prime z^\prime$ filter passbands,
after excising $\pm 1500$\,km\,s$^{-1}$ wide bands around the strong wind lines (mostly He~I, but also
Ca and Mg) seen at binary minimum (Figure \ref{EmLines}). To normalize these filter bandpasses
to standard SDSS magnitudes, we integrated the comparison-star spectra over the same bands. 
Carefully calibrated GROND observations give magnitudes for this star
of $g^\prime=18.58$, $r^\prime=18.05$, $i^\prime=17.79$, and $z^\prime=17.70$\,mag
(A. Rau, private communication). This determines an offset to the measured magnitudes 
for each spectrum. Comparing with synthesized magnitudes for the full SDSS passbands,
we do indeed find decreased source fluxes, by as much as a magnitude when the wind 
lines are strong.

\begin{figure}[t!!]
\vskip 9.6truecm
\includegraphics{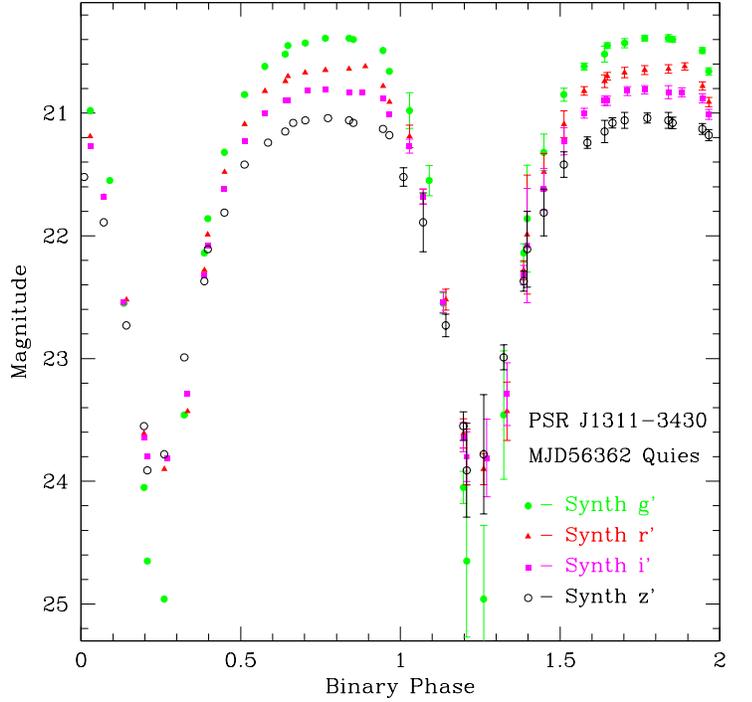}
\begin{center}
\caption{\label{SpecLCs} 
Synthesized band magnitudes from the Keck MJD 56362 spectra. These $g^\prime r^\prime i^\prime z^\prime$
magnitude estimates are formed by integrating over the SDSS passbands after excising
strong (mostly He~I) emission lines. Magnitudes are normalized by matching the synthesized 
comparison-star magnitudes to the direct imaging values. Two periods are shown with estimated errors during
the second cycle (see text).
}
\end{center}
\vskip -0.7truecm
\end{figure}

	However, even this line excision does not completely remove the light-curve 
fluctuations. To best estimate a ``quiescence'' light curve, we use our spectra
from MJD 56362, and compare the synthesized magnitude estimates in $\Delta \phi=0.05$
phase bins. We adopt the largest (faintest) magnitude in each band in each bin as the quiescence 
value and plot the resulting light curves in Figure \ref{SpecLCs}. Statistical
errors are obtained from the fluctuations in the spectroscopic fluxes integrated to derive
the band magnitudes. We further estimate the remnant light curve fluctuations associated 
with faint wind lines or continuum flares, by taking the difference to the second 
faintest magnitude observed for a given bin. These error estimates are added in 
quadrature to the statistical errors and plotted on the bin points in Figure \ref{SpecLCs}.

	Several features of the derived light curves deserve comment. First,
our synthesized magnitudes, normalized to the comparison star, are roughly 0.6\,mag
fainter than those reported from direct GROND imaging in R12. It is possible
that this apparent 40\% flux decrease is caused by differential slit losses between
the pulsar and comparison star. However, the GROND magnitudes at MJD 56118 may
have been well above quiescence, since these data show larger fluctuations (even 
at $\phi_B \approx 0.75$) than seen in these synthesized magnitudes. Of course,
even these new light curves will not represent complete quiescence --- fluctuations are still
visible across the maximum --- but they provide a substantially improved target
for quiescence modeling.
Overall the light curves follow the form of the direct imaging SOAR/GROND curves
reported by R12, although we appear to have better captured the red colors at binary
minimum.  It should be noted that as in R12, the light curves remain slightly asymmetric,
brighter at ``Dusk" ($\phi_B \approx 1.0$) by $\sim 0.1$\,mag compared
to Dawn ($\phi_B \approx 0.5$).

\begin{figure}[b!!]
\vskip 9.3truecm
\includegraphics{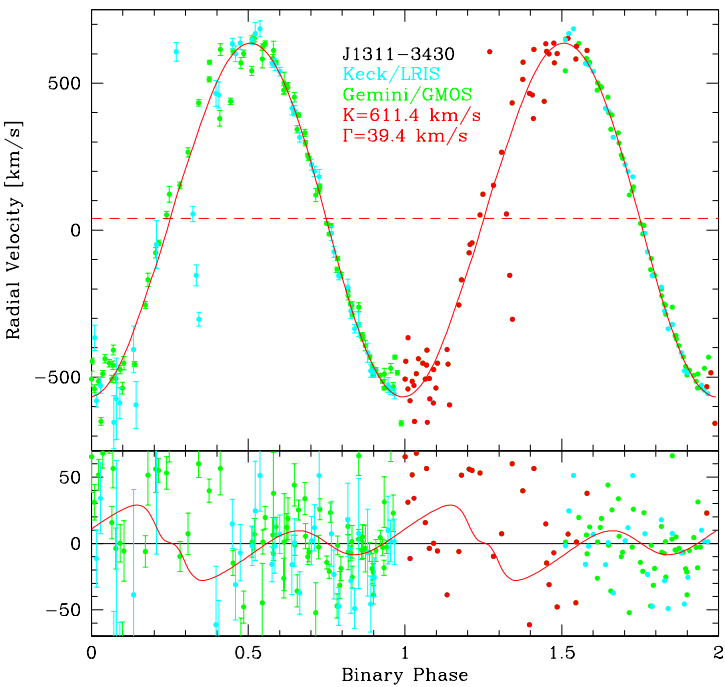}
\begin{center}
\caption{\label{RV} 
J1311 radial-velocity measurements from LRIS and GMOS-S data. The first cycle shows
all $R>1$ correlation points, with velocity error flags. During the second cycle
the data used in the simple sinusoidal fit (solid line, dashed line marks the systemic velocity $\Gamma$) 
are in green and cyan; red points are excluded.
The bottom panel show the residuals to this best-fit sinusoid. The curve 
plots residuals to the ELC velocity curve (``ELC-Sp" model) after subtracting the simple sinusoid,
demonstrating that the model's non-sinusoidal radial-velocity terms are present
in the data (see text).
}
\end{center}
\vskip -0.7truecm
\end{figure}

\section{The Companion Radial Velocity}

	The dramatic heating offsets the CoL in a poorly understood,
temperature-dependent way.
Also the variable wind emission in precisely the He~I lines dominating the companion
photosphere can shift the absorption-line centroids. These effects make measurement of
the companion radial velocity challenging. We focus here on the 
March 2013 Keck/LRIS and the May-June 2013 Gemini/GMOS-S datasets. 
For the Keck data we start by removing a scaled version of the emission-line 
velocity profiles (Figure \ref{EmLines}) from each detected line feature in
the blue channel. We then use a He-dominated model atmosphere template from
the atlas of \citet{jwp01} ($T_{\rm eff} =$ 12,000 $\pm$ 1000\,K, log\,$g=4.6\pm0.2$)
as a template for velocity cross-correlation to obtain the best radial 
velocity of each individual spectrum. Cross-correlation is performed using the 
IRAF RVSAO package \citep{km98}. 
The quality of the fit is monitored by the correlation statistic $R$ (the ratio 
of the correlation peak height to random peak height), which
varies from $\sim 12$ on the day side, where velocity uncertainties are
as small as $\sigma_v=15$\,km\,s$^{-1}$, to an insignificant $R<1$ on the Night side.
Measurement of the comparison star with a G-star template showed a slow
drift in radial velocity from +30\,km\,s$^{-1}$ to $-30$\,km\,s$^{-1}$, with typical
uncertainties of 6\,km\,s$^{-1}$. We accounted for this residual calibration drift
and also corrected to heliocentric velocities.

	Similar measurements were made of the GMOS-S spectra. Here we reach $R \approx 7$ 
during the Day phases. With the high resolution, this delivered velocity uncertainties 
as small as $\sigma_v\approx 5$\,km\,s$^{-1}$. Night-time cross correlations were often
insignificant, but occasionally delivered apparently significant $R \approx 1$--2;
some of these measurements were close to the orbital curve while others had
large departures. The comparison-star radial velocity was monitored to 
typically $\sigma_v \approx 4$--6\,km\,s$^{-1}$, although uncertainties grew to as large as 
$\sim 10$\,km\,s$^{-1}$
when conditions were poor. No strong velocity trend was evident in these data. 

	The corrected cross-correlation velocities for these two data sets are
show in Figure \ref{RV}. Two periods are illustrated, phased to the orbital solution
of \citet{pet12}. During the first, all points with $R>1$ are plotted, with their
$\sigma_v$ error flags. The radial-velocity curve during the Day phase is very well
measured. During the Night phase a number of GMOS-S points appear to follow the
expected curve, but many are displaced by $\sim +30$--60\,km\,s$^{-1}$; the few plotted LRIS
points in this phase range seem largely random. Recalling that the GMOS-S data measure
only the blue portion of the spectrum, it may be that the radial velocity is tracking flux 
from the bright lune of the heated side, visible at the companion limb for $i<90^\circ$.

	In addition to the low-statistics Night-time points, it is clear that there are
large variations from the radial-velocity curve at the Dawn and, especially, 
Dusk phases. At times the velocities follow the simple sinusoid, at others the
velocities lie up to $\sim 100$\,km\,s$^{-1}$ ``inward'' of the expected curve,
to the blue ($-v$) near $\phi_B=0.5$ and to the red ($+v$) near $\phi_B=0.0$.
These offsets are highly significant and do correlate among spectra. However,
we are not able to associate these with, say, periods of strong wind emission lines.
One interpretation is that these offsets represent radial outflow near the companion
terminator; an emission component to the red (blue) at Dawn (Dusk) would impose a variable
blueshift (redshift) on the absorption features measured against the cross-correlation
template. More high-quality data are needed to pin down the physical origin of
this effect.

	One consequence of the terminator shifts is a difficulty in fitting a simple
radial-velocity curve. If we fit only to correlation measurements with $R>3$,
and restrict to $\phi_B=0.5$--1.0, we obtain an observed radial-velocity
amplitude $K_{\rm CoL}=611.4 \pm 5.4$\,km\,s$^{-1}$ and systemic velocity
$\Gamma=39.4 \pm 3.7$\,km\,s$^{-1}$ (statistical errors only). The fit departures remain substantial, with a
$\chi^2$ per degree of freedom (DoF) of 9.8 for 74 DoFs. If we restrict further to the Day
side ($0.55<\phi_B<0.95$), we obtain $K_{\rm CoL}=615.4 \pm 5.1$\,km\,s$^{-1}$, with
$\chi^2$/DoF = 4.0. In contrast, fitting all points with $R>3$ includes many discrepant
Dusk points and gives $K_{\rm CoL}=590.7 \pm 7.2$\,km\,s$^{-1}$, with $\chi^2$/DoF = 15.6.
It is evident that non-sinusoidal terms are present. The bottom of Figure \ref{RV} shows the
residuals to the simple sinusoidal fit. Despite the scatter, some systematic trend
is visible, especially during the Day side. The curve plotted, for a
best-fit ELC model with an L1 cool region (see below), shows a similar
residual to the fit sinusoid, and the fit to this curve decreases $\chi^2$ by 53.7,
i.e. $\chi^2$/DoF decreases from 9.8 to 9.1.

\section{Models of the Heated Photosphere}

	There are a number of codes used to model light curves and radial velocities in
interacting binaries. We find that none is fully adequate to cover the extremes
of temperature, surface composition, and pulsar heating found in J1311. Nevertheless,
such modeling fails in instructive ways and provides a guide toward future, more detailed
solutions. We report here on modeling with the ELC code (Orosz \& Hauschildt 2000)
and the ICARUS code (Breton et al. 2011), both kindly made available to us by their authors.
For both we can run in ``MSP" mode, with the period $P_B=0.065115$\,d and pulsar orbit 
$x_1=a_1 {\rm sin}\,i = 0.010581$\,lt-s held fixed. Both codes model the effect of pulsar
heating of the companion as illumination by a point (X-ray) source.

\begin{figure}[t!!]
\vskip 9.4truecm
\includegraphics{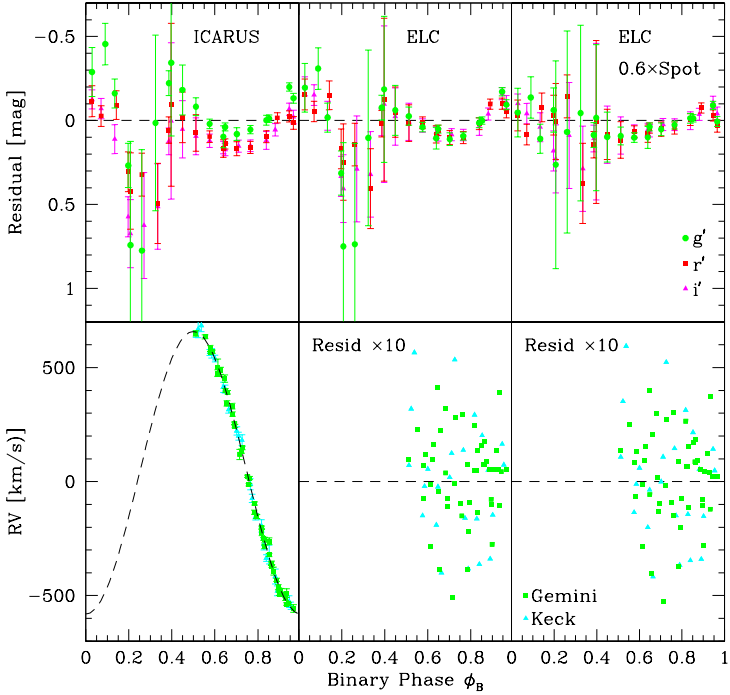}
\begin{center}
\caption{\label{fitresid} 
Residuals to the model fits. Top: photometric ($g^\prime r^\prime i^\prime$) residuals with
(left to right) the ICARUS fit, the ELC baseline fit, and the ELC fit with a $0.6 \times L_1$ spot.
Bottom left: the radial-velocity points used in the ELC fitting. Middle and right panels show the
residuals to the ELC baseline and $L_1$ spot fits, respectively.
}
\end{center}
\vskip -0.7truecm
\end{figure}

	For the ICARUS code we ran with color tables generated from the BTSettl atmospheres
(Allard et al. 2010). In all cases the best-fit models preferred large Roche-lobe fill
factors; we set the effective fill factor to 0.99. In fitting our quiescent 
$g^\prime r^\prime i^\prime z^\prime$ light curves, the main parameters are the
orbital inclination $i$, the Night side (unheated) star temperature $T_N$, and the heated
side (Day) temperature $T_D$. We report this last parameter as an effective heating luminosity
$L_H= 4\pi a^2 \sigma (T_D^4-T_N^4)$, where the orbital separation is $a\approx 0.95\, {\rm R}_\odot$
and the heating luminosity can be compared with the pulsar spin-down power
${\dot E} = 5 \times 10^{34}\, I_{45}\, {\rm erg \,s^{-1}}$.
ICARUS also fits the absolute fluxes and so returns
a distance estimate and an effective extinction. If left free, the extinction
fits to an insignificantly small value; to be conservative, we set the value to 
the low net Galactic extinction $A_V=0.173$\,mag in this direction (Schlafly \& Finkbeiner 2011).
The fit values are given in Table 2, with uncertainties as reported by the ICARUS fitter.
The distance is surprisingly large at 2.6\,kpc, and the heating flux slightly exceeds the 
standard pulsar spin-down power. Most importantly, the fit inclination is so small that with
the observed radial-velocity amplitude $K_{\rm CoL}=615$\,km\,s$^{-1}$, the minimum pulsar mass would be 
$2.77\,{\rm M}_\odot$. Since the true kinematic amplitude is larger 
(typically with $K_{\rm cor} > 1.03$, Table \ref{Fitvals}),  
then $K_{\rm CoM}> 633$\,km\,s$^{-1}$ and the inferred
mass would be unphysically large at $>3\, {\rm M}_\odot$.  Also, the fit residuals (Figure \ref{fitresid}) demonstrate 
that the maximum is substantially wider than the model light curve, and that the
colors are redder at peak than predicted by the model. Clearly, the present ICARUS model
is inadequate for a detailed fit.

\begin{deluxetable*}{lllllll}[h!!]
\tablecaption{\label{Fitvals} J1311$-$3430 Light Curve/spectral Fits}
\tablehead{
\colhead{Parameter} & \colhead{Units}&\colhead{ICARUS}& \colhead{ELC} & \colhead{ELC-Sp} & \colhead{ELC-Sp/$iq$}\cr
}
\startdata
Incl.           &Deg.            &$57.1\pm 1.5$  &$64.0\pm 3.5$        &$81^{+4b}_{-6}$        & $77^{+3}_{-4}$   \cr
$T_N$           &K               &$4240\pm 150$  &$4800\pm400$         &$5000\pm400$           & 4500$^c$ \cr
${{\rm Log}(L_H)}$&erg\,s$^{-1}$           &$34.80\pm 0.03$&$35.5_{-0.3}^{+0.15}$ &$35.6\pm0.2$           & 35.0$^c$   \cr
$q $            &                & --            &$184.5\pm 1$           &$179.5^{+1}_{-0.5}$   & $180\pm0.5$      \cr
$K_{\rm cor}$  &                & --            & 1.062               & 1.033                  & 1.036 \cr
$M_1^a$         &M$_\odot$       & --            & 2.63$_{-0.2}^{+0.3}$ &$1.82^{+0.15}_{-0.06}$ &$1.88^{+0.09}_{-0.06}$   \cr
$M_2^a$         &$10^{-2}\,{\rm M}_\odot$& --            & 1.42$\pm 0.15$      &$1.01^{+0.09}_{-0.02}$  &$1.04^{+0.05}_{-0.03}$ \cr 
$\chi^2$/DoF    &148             & --            & 3.74                & 2.60                   &2.66     \cr
\enddata
\tablenotetext{a}{Value inferred from fit -- projected uncertainties, errors inflated by $\chi^2$/DoF.}
\tablenotetext{b}{Bounded by lack of an eclipse.}
\tablenotetext{c}{Fixed in fit, with spot depth $0.6\times$.}
\end{deluxetable*}

	The ELC code was used to fit the light curves and spectra simultaneously. For these
fits we used an atmosphere model table computed for SDSS filters by J. Orosz, extracted from the ``NextGen" 
atmosphere grid with an extension to $T_{\rm eff} >$ 10,000\,K using \citet{ck04} models.
As before, the fits preferred large Roche-lobe fill factors and this value was fixed at 0.99.
We again fit for the inclination, $T_N$, and heating, but additionally constrained the
mass ratio $q=M_1/M_2$ for direct estimates of the component masses. The results, which include 
radial-velocity constraints, prefer larger inclinations and lower masses than the ICARUS solution.
The $T_N$ temperature is slightly higher, but most importantly the fit $L_H$ is much higher and is
indeed substantially larger than the total spin-down power.
Again, the observed maximum is appreciably wider than the heating model suggests, and the mid-Day ($\phi=0.25$)
temperature is cooler than predicted. The fit is not statistically satisfactory and implies a large pulsar mass
$m_1=2.6^{+0.3}_{-0.2}$. Here, and for all ELC fits, we quote ``$1\sigma$" errors as the full multi-dimensional
projection over all fit parameters of the region about the fit minimum with a fit statistic
increase of $\chi^2_{min}/{\rm DoF}=1$; in this sense our uncertainties are conservative, with marginalization
over all other parameters and error inflation for imperfections of the model.

	The ELC code has the option of applying artificially heated or cooled star spots to 
the secondary. As in R12, we can substantially improve the fit by cooling the
inner Lagrange point. We apply a spot which decreases the local unperturbed surface temperature
with a Gaussian profile of radius $45^\circ$. The best fits invoke a $\sim 40$\% temperature decrease 
and do give a substantially better model for the light-curve maximum.  The result is that the 
companion is brightest in a ring pulsar-ward of the terminator, with the CoL 
much closer to the CoM than for direct heating (small $K_{\rm cor}$). In addition, the heating
pattern broadens the light-curve maximum. This drives
the inclination $i$ to larger values, to maintain the depth of the light-curve minimum. Both
of these effects serve to decrease the inferred pulsar mass. 

\begin{figure}[t!!]
\vskip 9.4truecm
\includegraphics{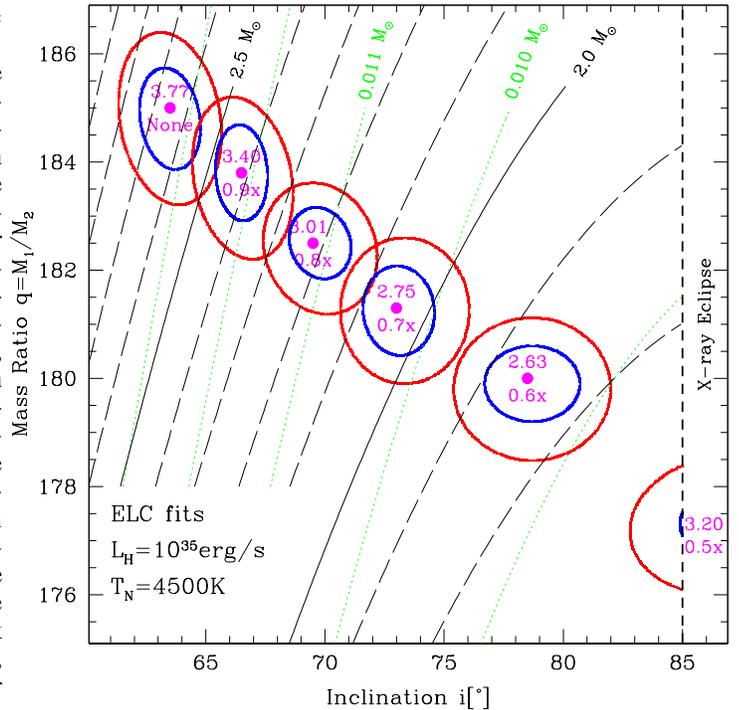}
\begin{center}
\caption{\label{spots} 
Confidence regions for two-parameter ELC model fits, holding $L_H$ and $T_N$ fixed near
their maximum values and varying the assumed spot deficit from $1.0\times$ (no spot)
to $0.5\times$ (deep spot, excluded by the lack of an eclipse). The best-fit point and
the $\chi^2$/DoF are marked for each deficit. The uncertainty ellipses are for $\chi_{\rm min}^2 + (\chi^2$/DoF)
and $\chi_{\rm min}^2 + 3\times (\chi^2$/DoF).  Lines of constant $M_1$ (solid, dashed) and $M_2$ 
(dotted) are drawn.
The inclination and mass ratio, and hence the inferred pulsar mass, depend dramatically on the 
heating pattern. 
}
\end{center}
\vskip -0.7truecm
\end{figure}

	While there are physical effects (see below) which can accommodate a larger effective
heating power, $L_H = 10^{35.6}\, {\rm erg\,s^{-1}}$ is much larger than the pulsar can supply.
Also, the large $T_N$ gives a poor match to the relatively red colors and large magnitude
at $\phi_B= 0.25$. The drive to these large values is a consequence of the
spot approximation for the surface-temperature distribution. If we fix $L_H = 10^{35}\, {\rm erg\,s^{-1}}$
and $T_N=4500$\,K, the largest values consistent with the observations, we find
that the inferred inclination (and thus neutron-star mass) remain a very strong function
of the spot temperature decrement (Figure \ref{spots}), which is rather poorly
determined. The best fit, with a  $0.6\times$ spot (40\% spot decrement), implies $M_1 = 1.9\,{\rm M}_\odot$. 
Weaker spots lead to large inferred masses, stronger spots require edge-on orbits in conflict 
with the lack of an X-ray eclipse \citep{r12}.  Thus, our conclusions depend sensitively on details of the heating model.
Notice in Figure \ref{fitresid} that the radial-velocity residuals vary only weakly between 
quite different heating patterns. The main effect is an overall increase in amplitude,
which is reflected in the changed $K_{\rm cor}$; the temperature (color) is more sensitive to the
heating model.

\section{Conclusions: Indirect Heating, Neutron-Star Mass}

	Our goal of measuring the orbital inclination and the $K_{\rm cor}$ factor
with light-curve fitting has been frustrated by the poor fit to the light-curve
maximum. A strong clue to the nature of the difficulty is that simple direct-heating fits require
a heating power larger than can be supplied by the $5 \times 10^{34}\, I_{45}\, {\rm erg\,s^{-1}}$
spin-down power.
For the ICARUS model, $L_X$ exceeds the nominal spin-down power by a factor of 1.3.
For the basic ELC model, the excess is a factor of 6, and for the ELC spot model, the heating power
exceeds that supplied by the pulsar by a factor of 8. This is especially worrisome as the typical
heating efficiency estimated by \citet{bet13} is $\eta=0.15$ of the pulsar spin-down.

	This discrepancy is so large that a number of factors are likely relevant. First,
the effective moment of inertia and resulting spin-down power may be larger than assumed,
especially if $M_1$ is $>2\, {\rm M}_\odot$ where $I_{45} \approx 2$--4, depending on the equation of state
\citep{ls05}.
Also, the present modeling assumes normal composition in the atmosphere; He-dominated spectra
can be bluer, allowing a lower fit $T_{\rm eff}$.
Most importantly, the assumption of direct, radiative heating in the ICARUS and ELC models is likely
inadequate for J1311. Instead, we expect that the pulsar wind terminates at an intrabinary shock 
wrapped around the $L_1$ point, reprocessing the pulsar wind into high-energy particles which deliver
heat below the companion photosphere. This reprocessing can both increase the
effective heating flux over direct radiation, as the intrabinary shock subtends a larger angle
at the pulsar, and preferentially heats the companion terminator. Further, Coriolis forces break the shock 
symmetry in the rotating frame and may thus provide a mechanism for explaining the asymmetric 
light-curve maximum.  The presence of strong companion fields at the $L_1$ point (indicated by the large flares)
may also preferentially direct intrabinary shock particles away from $L_1$. 
Such effects are dramatic for J1311, but may be present at lower levels in other evaporating pulsar binaries,
where high-precision photometry shows asymmetric light curves \citep[e.g.,][]{sh14}.
This may bias mass estimates inferred from fitting light curves with simple direct-heating models.

	Our spectroscopic campaign has revealed a number of new aspects of this remarkable system.
First, J1311 undergoes violent flaring in the optical, which suggest strong magnetic activity on
the heated face of the companion. Second, strong (but intermittent) line emission is generated
in the wind driven off the companion. This outflow spectrum is dominated by neutral helium lines
and appears to follow preferred paths corotating with the binary system.  Since the line flux along
these paths is sporadic, further study is needed to determine the persistent patterns, but
emission along a pulsar wind-driven spiral outflow provides a viable model. Selecting against
companion flares and spectrally excluding the outflow line flux, we have made improved measurements
of the companion photosphere's heated light curve and color variation. These data strengthen the conclusion that 
simple radiative heating models are not adequate to describe strongly interacting black-widow
binaries like J1311. Instead, we posit that the heating flux is reprocessed in an intrabinary
shock whose illumination of the companion surface preferentially heats away from L1, toward 
the Day-Night terminator. If this deep heating is mediated by shock-accelerated particles,
coherent magnetic fields near L1 can also help redirect the flux and decrease radiation from the L1 point.

	The net effect of asymmetric heating is to prefer smaller $K_{\rm cor}$ and larger sin\,$i$, but the
precise values of these critical parameters are very sensitive to the heating pattern. There
are two important paths to improving their determination. (1) Improved, more physical, modeling
can better exploit existing measurements. However, statistically acceptable $\chi^2$ will probably
require excellent exclusion of flare perturbations, even if the heating model is physically correct.
(2) Observationally, one would like to directly measure the heating pattern across the companion Day side. 
Improved spectroscopy remains the key. While well-measured colors, especially near minimum ($\phi_B=0.75$),
can help constrain the fits, we have seen in this study that these can best be obtained by synthesizing
colors from spectra, after excluding flares and the intermittent wind emission lines. Of course,
with sufficient S/N, the spectra themselves will help determine the $T_{\rm eff}$ distribution
as a function of phase, by measuring relative absorption-line strengths for species with differing excitation.
Even more directly, higher spectral resolution with good S/N can measure rotational
broadening of the companion's absorption lines and its variation with orbital phase, 
thereby providing perhaps the cleanest constraints on sin\,$i$.

	Until such improvements are made, we cannot exclude any mass in the range 1.8--2.7\,M$_\odot$
for PSR J1311$-$3430. This range of values still allows a wide range of system evolutions and may,
or may not, constrain the equation of state at high densities. Thus, more work is needed to fully exploit
the promise of this unusual system.

\bigskip

\acknowledgements

We thank Kelsey Clubb, Ori Fox, and Melissa Graham for assistance with some of the Keck observations, 
as well as German Gimeno for help with the Gemini campaign. We also thank the referee
for many detailed comments.
This work was partially supported by NASA grants NNX11AO44G and NNX12A068G. 
A.V.F. and S.B.C. were supported by Gary and
Cynthia Bengier, the Richard and Rhoda Goldman Fund, the Christopher
R. Redlich Fund, the TABASGO Foundation, and NSF grant AST-1211916.  
Some of the data presented herein were obtained at
the W. M. Keck Observatory, which is operated as a scientific
partnership among the California Institute of Technology, the
University of California, and NASA; the Observatory was made possible
by the generous financial support of the W. M. Keck Foundation.


\end{document}